\documentclass[a4paper,fleqn,usenatbib,useAMS]{mnras}

\usepackage{graphicx}	
\usepackage{amsmath}	
\usepackage{amssymb}	
\usepackage{multicol}        
\usepackage{bm}		
\usepackage{pdflscape}	

\usepackage[T1]{fontenc}
\usepackage{ae,aecompl}

\title{The Dynamical Connection Between Phaethon and  Pallas}

\author[N. Todorovi\' c]{
Nata\v sa Todorovi\' c \thanks{E-mail:ntodorovic@aob.rs}
\\
Belgrade Astronomical Observatory, Volgina 7, P.O.Box 74 11060 Belgrade, Serbia
}

\date{Accepted XXX. Received YYY; in original form ZZZ}

\pubyear{2018}

\begin{document}
\label{firstpage}
\pagerange{\pageref{firstpage}--\pageref{lastpage}}
\maketitle

\begin{abstract}
Phaethon (3200) is an active Near Earth Asteroid classified as a B-type object and a suspected former member of the Pallas family. 
In this article, we search for sources of Phaethon-like orbits originating from the two strongest resonances in the region of  the Pallas family; namely, 
the 5:2 and 8:3 mean motion resonances with Jupiter, located at $a\sim 2.82$ AU and $a \sim 2.70$ AU, respectively. The probability for this dynamical 
connection observed in \cite{deLeon2010} was close to $2$ per cent. Using sophisticated numerical methods, we found a highly efficient dynamical flow between 
Pallas and Phaethon. As many as  $43.6$ per cent of test particles placed in the 5:2 resonance were able to reach the orbit of Phaethon; whereas, 
for the 8:3 resonance, this percentage was $46.9$ per cent.
\end{abstract}





\section{Introduction}

The Near Earth Asteroid (3200) Phaethon  was found  by the Infrared Astronomical Satellite \citep{GreenKowa1983} in October, 1983 
and it was identified as the long sought parent body of the Geminid meteor shower \citep{Whipple1983, Gustafson1989, Williams1993}.
Not long after,  it turned out that  unlike other meteoroid streams originating from comets, Phaethon, the parent body of the Geminids - is an asteroid. 

The activity  of this unusual asteroid  has been largely studied in the literature. According to
\cite{Delbo2014} the large eccentricity of Phaethon's orbit ($e\sim 0.89$) plays an important role in its  geological activity. 
Periodically approaching to and then distancing from the Sun 
causes cyclical cooling and heating, leading to the thermal cracking of its mineralogical surface.
The intensive solar radiation near perihelion at $q\sim 0.14$ AU
ejects away small grain-sized particles that enter into the Earth's atmosphere producing Geminids. 
However, one of the open questions in this scenario is that the estimated mass loss from Phaethon $\sim 3\cdot 10^5$kg \citep{Jewitt2013}
does not account for the total mass of the Geminids $ \sim 10^{12}-10^{13}$ kg \citep{Jenniskens1994}.

Spectral attributes of  Phaethon (see for example \cite{Hanus2016} and references therein) classified this asteroid as a B-type object \citep{DeMeo2009}. 
The largest population of such bodies can be found in the asteroid family Pallas, indicating that Pallas could be the dynamical origin of Phaethon. 
Also, there should be an efficient dynamical mechanism able to drive bodies from the part of the main belt where the 
Pallas family is located, down to the highly inclined near-Earth region of Phaethon. 
In \cite{deLeon2010}, it was illustrated that Phaethon is connected both spectroscopically and dynamically to the Pallas family, and that the major role in this transportation is played by the two strongest 
mean motion resonances (MMRs) with Jupiter in the region of Pallas family: the 8:3 at $a\sim 2.70$ AU and the 5:2 at $a\sim 2.82$ AU.  However, a relatively low fraction of test particles (only $\sim 2\%$)  
placed into the two resonances where actually able to recover a Phaethon-like orbit.

Our goal is to revisit the role of the 8:3 and 5:2 MMRs in the Pallas-Phaethon dynamical connection. 
Using sophisticated numerical methods, 
we illustrate that this dynamical link acts with a much higher efficiency than hitherto observed.

\section{Method and Results}
\label{maps}
 
The standard way to study the dynamics of a resonance is to inject into it a certain number of test objects, 
clone the particles (with the corresponding standard deviation) using Gaussian distributions in the orbital element space, 
and integrate them in some fixed time, during which we observe and study their dynamical evolution. 
In using this method, however, it is very common that only a small fraction of test particles interact with the resonance, despite long computational times.

Here we study the resonant dynamics using  a method  introduced in  \citet{Todo2017}. The first step is to calculate short-term dynamical maps of the resonance 
using the Fast Lyapunov Indicator -
FLI \citep{FLGonz1997, FGonzL1997}. It is important that the calculation times of the maps are short, because in this way the fine structure of chaos in the 
resonance can be clearly detected \citep{GuzzoLega2014, Todo2017}.
The more unstable parts, easily noticeable on the maps, represent 'good' starting positions for diffusion.
Particles chosen along them should be active as soon the integration starts, making this method computationally cheaper and more efficient. 

Let us mention that originally, FLI was used for idealized systems such as symplectic maps or simplified Hamiltonians \citep{Froeschle2000Science, LGFArndiff2003, FGLLocGlobdiff2005, Todo2008, Todo2011}, 
but later it was successfully applied to studies  of asteroids, planetary and satellite systems as well, 
from the close neighborhood of Earth \citep{Daquin2016, Rosengren2015} to the outer Solar System \citep{Guzzo2005, Guzzo2006}
or exoworlds (see for example \citet{PilatDvorak2002, Dvorak2003, Sandor.etall2007, Schwarz2011}).

\subsection{The 5:2 resonance}
\label{sub52}

\begin{figure}
\includegraphics[width=0.57\textwidth]{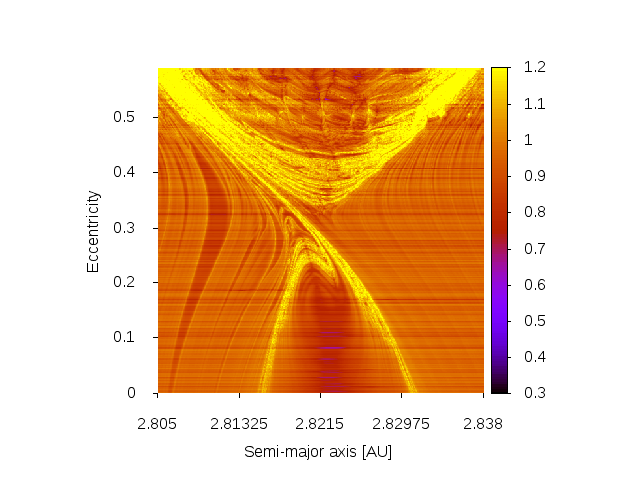} 
\caption{The 5:2 mean motion resonance with Jupiter computed in the orbital plane of Pallas, for 5 Kyrs and  for $ a\times e = [2.805,2.838]\times[0,0.589]$. 
Other orbital elements are fixed to $i=34.80 \deg$, $\Omega=173.09$, $\omega=310$ and $M=248.97$. 
Stable particles are red, while the most chaotic ones are yellow. The $(a,e)$ profile of the resonance has an atypical hourglass structure and 
the location of the unstable point is unusually high (close to $e\sim 0.3$).  }
\label{FIG1}
\end{figure}

Since we are interested in the Pallas-Phaethon transportation route, we calculate the FLI map of the 5:2 MMR with Jupiter in the orbital plane of Pallas,
that is for  $i=34.84\deg$, $\Omega=173\deg$, $\omega=310\deg$ and $M=249\deg$, using  $500\times 500$ test particles with semi-major axis and  eccentricity    in between 
$ a\times e = [2.805,2.838]\times[0,0.589]$. This map, computed for 5Kyrs, is given in Figure~\ref{FIG1}. The color scale, adjusted to FLI values is chosen such that  stable particles with $FLI<0.9$ are red, 
while the more chaotic ones with $FLIs$ larger than 1.1 are yellow. 

All our computations are performed using the ORBIT9 integrator\footnote{Available from \url{http://adams.dm.unipi.it/orbfit/}.} and 
the calculation time for one map is around 15 minutes\footnote{We use the Fermi cluster at the 
Astronomical Observatory of Belgrade. The cluster has 12 worker nodes, each node (HP SL390S blade server G7 X5675) has a 2xX5675(2x6core) processor, 
3.1GHz, 24G memory, 2Tb disc and 2xM2090 NVIDIA Tesla-fermi GPU cards.}. More technical details about the computations of the maps can be found in \cite{Todo2017}.

\begin{figure}
\centering\includegraphics[width=0.5\textwidth, height=0.27\textheight]{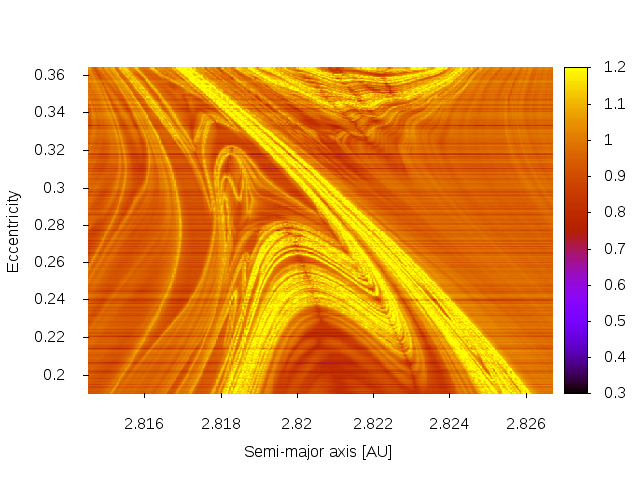}
\caption{The enlarged portion of the 5:2 MMR in the vicinity of the unstable point, where we can notice the peculiar deviation of the separatrix. 
Such shapes look very similar to the traces of the normally hyperbolic invariant manifolds. 
The candidates for diffusion are selected along the yellow structures. The range in eccentricities 
$e \in [0.19,0.364]$ includes the eccentricity of Pallas and other family members of the Pallas family.}
\label{FIG2}
\end{figure}

The $(a,e)$ resonant profile visible in Fig.~\ref{FIG1} has an hourglass structure, different from the typical V shape of a mean motion resonance. 
Other high-order resonances are also visible herein, but unlike the usual straight vertical V shaped resonant forms, those resonances are also deviated in the region of the unstable point. 
The unstable point (narrowest part of the resonance) is displaced to large eccentricity, close to $e \sim 0.3$, where we can notice a peculiar deviation of the separatrix. 

For a better illustration, an enlarged portion of Fig.~\ref{FIG1} close to the unstable point, in the region $ [a,e]=[2.814, 2.827]\times[0.190,0.364]$ is plotted on the Fig.\ref{FIG2}. 
The deviated shapes of the separatrix remind strongly on the so-called hyperbolic invariant manifolds of the saddle point of the resonance \citep{morbidelli2002modern}, the crucial elements in the production of chaos.
Let us mention that invariant manifolds are identified as sources of natural transportation  routes, and are therefore very important in space trajectory design and astrodynamics 
in general   \citep{Gomez2004,  Vilac2008, Bello2010}. 
Their localization  however, is not trivial, even in highly restricted dynamics. Algorithms for their numerical evaluation can be found for example in \citet{Simo1999, Koon2011, Topputo2016}.
Treating a circular restricted three-body problem, \cite{GuzzoLega2014} illustrated that invariant manifolds can be captured also by short-term FLI maps.
Using this ability of FLI and a real  Solar System  model \footnote {We treat all planets from Venus to Neptune, while the 
mass of Mercury is added to the mass of the Sun and the corresponding barycentric correction to initial conditions is applied.}  in \cite{Todo2017} we detected hyperbolic manifolds-like  structures in the 5:2 MMR with Jupiter, and we illustrated that such structures 
are fruitful sources of fast diffusing orbits. We have also observed that  8 per cent  of test bodies  placed into the 5:2 MMR recovered a Phaeton like orbit.

Following the same methodology as in  \cite{Todo2017}, but focusing on the dynamical pathways between Phaethon and Pallas, we change the initial orbital plane inside the 5:2 MMR, as described above. 
Then, we chose 1000 of the most chaotic particles in between $e \in [0.19,0.364]$ along the intricate yellow structures on the Fig. \ref{FIG2}.  
An eventual injection of Pallas fragments into the 5:2 MMR fits well in the chosen range in eccentricity, since  the eccentricity of Pallas is $\sim 0.23$ 
and the remaining Pallas family members have $0.255 < e < 0.283$. 
We follow their  orbital evolution for 5 Myrs and count all the bodies that at some moment  recovered the orbit of Phaethon. 
That is, we search in the $(a, e, i)$  element space those particles that satisfy 
\begin{equation}
 |a - a_{Ph} | < 0.1, |e - e_{Ph}| < 0.1,  |i - i_{Ph} | < 3,
 \label{pha}
\end{equation}
where $ (a_{Ph}, e_{Ph}, i_{Ph})$ are the respective semi-major axis, eccentricity and inclination of Phaethon. 
These elements are taken from the AstDys database\footnote{Available at \url{http://hamilton.dm.unipi.it/astdys/}.} 
and their  osculating  values for the epoch 57600.0 MJD are  $a_{Ph}=1.271$ AU, $e_{Ph}= 0.889$ and $i_{Ph}=22.24 \deg$.

Mapping the resonance and selecting particles in the  orbital plane of Pallas, strongly affected the number of bodies reaching Phaeton.
Instead of 8 per cent observed in \cite{Todo2017}, we found that among  1000 test bodies placed into the 5:2 MMR,  436 objects 
satisfy the condition (\ref{pha}). The first fragments placed into the resonance arrived to the current position of Phaethon after 0.294 Myrs
and their median delivery time is 1.7 Myrs.

\subsection{The 8:3 resonance}
\label{sub83}

\begin{figure}
\centering\includegraphics[width=.57\textwidth]{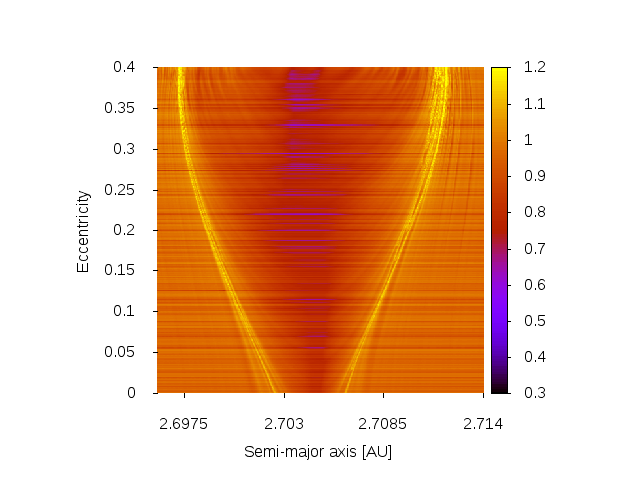}
\caption{The FLI map of the 8:3 mean motion resonance with Jupiter computed for 5Kyrs in the range
 $[a,e] =[2.696, 2.714] \times[0 ,0.4]$. Other numerical parameters are the same as for Fig. \ref{FIG1}.
 The particles for diffusion, candidates for reaching the orbit of Phaethon are selected along the resonant border in between  $ e \in [0.2 ,0.36]$.  }
\label{FIG3}    
\end{figure}

The map of the 8:3 MMR computed in the orbital plane of Pallas is illustrated in Figure~\ref{FIG3}. 
The FLIs are calculated for 5 Kyrs and  for $[a,e] =[2.696, 2.714] \times[0, 0.4]$, while the other orbital and numerical parameters are the same as those for Fig.~\ref{FIG1}.
Contrary from the 5:2 MMR, this resonance does have the standard V shape, the map is  depleted from peculiar structures, and chaos manifests only along the resonant border. 
Neither the unstable point or high-order neighboring resonances are visible on the map.

Again searching for Phaethon-like orbits,  we selected 1000 particles along the separatrix in the eccentricity range $ e \in [0.2 ,0.36]$ and 
repeated the same procedure as described above: we integrate the objects for 5 Myrs and count all the particles entering into the domain given in (\ref{pha}).

The 8:3 MMR delivered slightly more bodies to the current position of Phaethon than the 5:2 MMR,  in total 469, with  the first arrivals occurring after 0.53 Myrs.
The  median transportation  time along this route is 1.5 Myrs.

\begin{figure}
\centering\includegraphics[width=.5\textwidth]{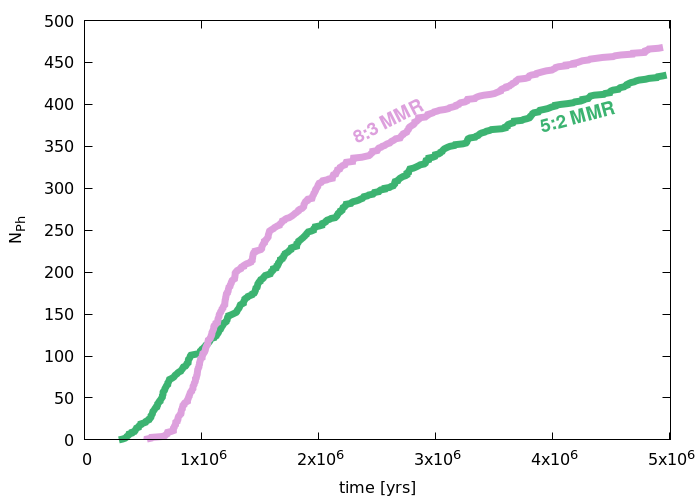}
\caption{The green line shows the number of  particles migrating  from the 5:2 MMR 
into the neighborhood of Phaethon and the pink line gives the same  for the 8:3 MMR.
Both resonances show similar transportation abilities.  The  5:2 MMR delivered 436 bodies to the neighborhood of Phaethon, and the 8:3 MMR was a little more efficient in that respect with 469 objects.
The first arrivals to the destination are detected after 294 500 and 532 500 years, respectively. }
\label{FIG4}
\end{figure}

Figure~\ref{FIG4} shows the influx rates of particles originating from the 5:2 (green) and 8:3  MMRs (pink).
In the first million years, the 5:2 MMR delivered more particles  towards Phaethon; yet, in the following Myrs the  8:3 MMR took over the transport role.
With $43.6$ and $46.9$ per cent of test bodies  delivered to Phaethon, the two resonances show similar transfer abilities, a result different from the one obtained in 
\cite{deLeon2010}, who found that  the  8:3 MMR was three times more efficient in this respect than the 5:2 resonance. However,  the result
in \cite{deLeon2010} was based on a smaller sample (only $2$ per cent) and the integration times were much longer, 100 Myrs.

Both Phaethon  and Pallas, source and destination region of our test particles, are in highly inclined orbital planes,  
which certainly affects the amount of material delivered along the route. 

\section{Conclusion}
\label{conclusions}

We have shown that the 5:2 and 8:3 MMRs with Jupiter possess a very powerful 
mechanism for transporting objects to the near-Earth region close to the asteroid Phaethon. 
The  results we obtained  are significantly different than earlier studies. 
In the work of \citet{deLeon2010},
where both 5:2 and 8:3 MMRs where studied the probability for this connection was  2 per cent, and in   \cite{Bottke.et.al.2002} for example, this probability was zero. 
In \citet{Todo2017} we applied the same methodology as done here and found that 8 per cent of test objects placed into the 
5:2 MMR reached the orbit of Phaethon; much less than in this work, because, in \cite{Todo2017}, the resonance was mapped in the lower inclined orbital plane of Ceres (at $i=10 \deg$). 
Choosing particles at the most unstable parts of the two resonances, but in the orbital plane of Pallas
increased the Pallas-Phaethon transportation efficiency multiply times.
Namely, almost half of test objects placed in the two resonances migrated to the neighborhood of Phaethon, 
strongly supporting the hypotheses of the Pallas origin of Phaethon and the Geminids meteor showers.

\section*{Acknowledgements}
This research was supported by the Ministry of Education, Science and Technological Development 
of the Republic of Serbia, under the project 176011 'Dynamics and kinematics of celestial 
bodies and systems'. The calculations were performed on a Fermi cluster located 
at the Astronomical Observatory of Belgrade, purchased by the project
III44002 'Astroinformatics: Application of IT in astronomy and close fields'.
The author is  grateful to A. Rosengren and the two reviewers  for useful remarks. 




\bibliographystyle{mnras}


\label{lastpage}
\end{document}